\documentstyle[tighten,preprint,pra,aps]{revtex}
\begin{document}
\draft
\title{All-electron Dirac-Coulomb and RECP calculations of excitation
energies
       for mercury atom with combined CI/MBPT2 method.}
\author{N.\ S.\ Mosyagin\cite{Email}, M.\ G.\ Kozlov, and A.\ V.\ Titov}

\address{Petersburg Nuclear Physics Institute,
         Gatchina, Petersburg district 188350, RUSSIA}
\date{\today}
\maketitle

\begin{abstract}
Calculations of transition energies between low-lying states of mercury
atom are performed in the frame of combined CI/MBPT2 method.
Results of all-electron relativistic calculations (using the
Dirac-Coulomb
Hamiltonian) are compared with experimental data and results of other
four-component
calculations.  The results of the RECP calculations are compared with
the
corresponding all-electron results in order to estimate accuracy of
different RECPs. Contributions from correlations in different
shells to the calculated excitation energies as well as effects of
basis set truncation at different orbital angular momenta, nuclear
models,
errors in gaussian approximation of the GRECP components are reported.
Analysis of the obtained results shows that at least 34 external
electrons
of mercury atom should be correlated and the one-electron basis set
should
contain up to $h$ angular momentum functions
in order to
reach a reliable agreement with experimental data within 200 cm$^{-1}$.
It is concluded that correlations of the $4f$ electrons can be
efficiently
taken into account for 20 electron GRECP at the generation stage.
\end{abstract}

\vspace{1cm}
{\bf SHORT NAME:} CI/MBPT2 calculations of mercury.

\vspace{1cm}
{\bf KEYWORDS FOR INDEXING:} Configuration Interaction, Many Body
Perturbation
 Theory, correlation structure (electronic structure), excitation
energies
 (transition energies), Relativistic Effective Core Potential
 (pseudopotential), four-component calculations, mercury,
 heavy atoms.
\pagebreak

\section{Introduction}

During the last few years, a considerable number of papers devoted to
electronic structure calculations of heavy atoms have been published
(e.g., see~\cite{Papers,Dzuba,Eliav}) that is not only due to
experimental
requirements but because they are good test systems to check
or estimate accuracy of different approximations before using them in
more
expensive molecular calculations.

In paper~\cite{Mosyagin}, Generalized Relativistic Effective Core
Potentials (GRECPs) were tabulated for atoms Hg through Rn and were
tested
in numerical Hartree-Fock (HF) calculations by comparison with the
all-electron Dirac-Fock (DF) and other RECP ones.  However, the question

of quality of the GRECPs for description of correlation effects
was not clarified yet. Whereas paper~\cite{Titov} contains an answer on
this question from theoretical point of view, the present paper is
devoted to our correlation structure calculations for mercury atom. For
this atom, the RECPs were generated by other groups~\cite{Ross,Hausser}
where the same number of electrons (20) was explicitly included into
calculations as in the case of the GRECP~\cite{Mosyagin}. Therefore, one

should expect about the same computational expenses in calculations with

all these RECPs and the comparison of their accuracy appears to be of
practical interest.

\section{Methods and basis sets}

The GRECP method was described in details in
papers~\cite{Tupitsyn,Mosyagin,Titov}. The main distinguishing features
of
this method are presence of non-local terms with the projectors on the
outercore (OC) pseudospinors together with the standard semi-local ones
in
the effective potential operator and generation of the effective
potential
components for smoothed pseudospinors which may have
nodes~\cite{Titov1}.

Theory of the CI/MBPT2 method is presented in papers~\cite{Dzuba}. In
this
method, the correlations in the valence (V) region of an atom which are
the most important ones are treated by the Configuration Interaction
(CI)
method (which is able to provide excellent results for the small number
of
correlated electrons) whereas the relatively small contributions to the
considered low excitation energies from the large number of the
core-valence correlations are taken into account with the help of
less expensive second order Many Body Perturbation Theory (MBPT2).

The program (in the $jj$-coupling scheme) for all-electron relativistic
four-component CI/MBPT2 calculations (using the Dirac-Coulomb
Hamiltonian)
was modified to make possible two-component RECP calculations (with the
nonrelativistic kinetic energy operator and relativistic $j$-dependent
potentials). This program allows one to use two different basis sets of
numerical spinors at the CI and MBPT2 calculation stages for
description of the correlations taking into account a space separation
of
the core and valence regions. Basis functions for the present
calculations
were obtained from numerical SCF (DF or HF) calculations of the
corresponding spinors for positive ion states. For example, the
[4,4,3,2,1] basis set for the 2 electron valence CI (2e-CI) was derived
from the calculations for the following nonrelativistically averaged
configurations of Hg:
$5d^{10} 6s^2$,        $[5d^9] 7s^1$, $[5d^6] 8s^1$, $[5d^3] 9s^1$,
$[5d^{10} 6s^1] 6p^1$, $[5d^9] 7p^1$, $[5d^8] 8p^1$, $[5d^7] 9p^1$,
$[5d^9] 6d^1$,         $[5d^6] 7d^1$, $[5d^3] 8d^1$,
$[5d^8] 5f^1$,         $[5d^7] 6f^1$,
$[5d^4] 5g^1$,
where the shells in the square brackets were frozen in the calculations
and the $1s$--$5p$ shells are dropped out for brevity.  For convenience
of
comparison, a $[k_s,k_p,k_d,k_f,k_g,k_h]$ basis set for the case of
$N$ correlated electrons will refer here to a basis set consisting of
$n_s s_{1/2},\allowbreak \ldots\allowbreak (n_s+k_s-1) s_{1/2};
n_p p_{1/2},\allowbreak \ldots\allowbreak (n_p+k_p-1) p_{1/2};
n_p p_{3/2},\allowbreak \ldots\allowbreak (n_p+k_p-1) p_{3/2};
n_d d_{3/2},\allowbreak \ldots\allowbreak (n_d+k_d-1) d_{3/2};
n_d d_{5/2},\allowbreak \ldots\allowbreak (n_d+k_d-1) d_{5/2};
n_f f_{5/2},\allowbreak \ldots\allowbreak (n_f+k_f-1) f_{5/2};
n_f f_{7/2},\allowbreak \ldots\allowbreak (n_f+k_f-1) f_{7/2};
5g_{7/2},\allowbreak \ldots (4+k_g)\allowbreak g_{7/2};
5g_{9/2},\allowbreak \ldots (4+k_g)\allowbreak g_{9/2};
6h_{9/2},\allowbreak \ldots (5+k_h)\allowbreak h_{9/2};
6h_{11/2},\allowbreak \ldots (5+k_h)\allowbreak h_{11/2}$
spinors where $n_s=5$ if $N\geq 20$, $n_p=5$ if $N\geq 18$, $n_d=5$ if
$N\geq 12$, $n_f=4$ if $N\geq 34$, otherwise $n_s=6$, $n_p=6$, $n_d=6$,
$n_f=5$.  Additional information on the used basis sets can be found on
http://www.qchem.pnpi.spb.ru. The above listed basis set has provided
the
lowest total energy in our CI calculations as compared with other tested

similar-sized basis sets.  More important advantage of this basis set is
a
possibility of one-to-one comparison between results of the all-electron

and RECP calculations not only for large (close to full) basis set sizes

but for the small ones as well. In fact, the calculated RECP errors
depend
only slightly on the size of such basis sets.

\section{Results and discussion}

Series of the correlation structure calculations were performed for the
number of correlated electrons varied from 2 to 34 and one-electron
basis
sets truncated at the orbital quantum numbers from 2 to 5.  The
all-electron relativistic calculations were implemented for two nuclear
models: a point nucleus and an uniformly charged ball with $1.334\cdot
10^{-4}$~a.u.\ radius. The GRECP calculations were made with both the
numerical GRECP components and their gaussian expansions
from~\cite{Mosyagin}. Moreover, 34 electron GRECP variant (where the
outercore $4f,\allowbreak 5s,\allowbreak 5p,\allowbreak 5d$ and valence
$6s,\allowbreak 6p$ shells of mercury atom are explicitly treated in
calculations) was generated and tested in these calculations.

The results of the 2e-CI calculations (where all the possible
excitations
of 2 valence electrons were considered and the
core $1s_{1/2}$--$5d_{5/2}$ spinors were frozen from the SCF
calculations
of the ground state with the $6s^2$ configuration) are presented in
tables~\ref{CI_large} and \ref{CI_small} for $[8,8,7,6,5,4]$
and $[4,4,3,2,1]$ basis sets. One can see from these tables that the use

of the $[4,4,3,2,1]$ basis set allows one to describe adequately the V-V

correlations and that the RECP errors are rather stable in respect to
the basis
set size variation.  Whereas the correlations only in the valence region
are
considered, the 20 electron GRECP is about 8 times more accurate than
the
20 electron RECP of Ross {\it et al.}~\cite{Ross} and about 30 times
more
accurate than the 20 electron energy-adjusted PseudoPotential
(PP)~\cite{Hausser} (first of all, due to the neglect of the difference
between the
outercore and valence potentials in the RECPs~\cite{Ross,Hausser};
see~\cite{Tupitsyn,Titov} for details).

From tables~III and IV in~\cite{Mosyagin} one can see that the errors
of the energy-adjusted PP, Ross {\it et al.}'s RECP, and the GRECP
for the excitations from the $6s_{1/2}^2$ state to
the $6s_{1/2}^1 6p_{1/2}^1 (J=0)$ and $6s_{1/2}^1 6p_{3/2}^1 (J=2)$
states\footnote{
Analysis for the case of the $6s_{1/2}^1 6p_{1/2}^1 (J=1)$ and
$6s_{1/2}^1 6p_{3/2}^1 (J=1)$ SCF states is more complicated
because they are strongly mixed in the CI calculations.
}
are -667 and +347~cm$^{-1}$, +182 and +224~cm$^{-1}$, -9 and
-9~cm$^{-1}$,
correspondingly.  They are in agreement with the errors from
tables~\ref{CI_large}
and \ref{CI_small} here with an exception of the GRECP errors (because
the
main contribution to the GRECP errors is due to errors of reproducting
the two-electron integrals rather than drawbacks of the effective
potential
operator).

In tables~\ref{CI_small}--\ref{MBPT_34}, results of both the
all-electron
relativistic and 20 electron GRECP calculations are presented for
different numbers of correlated electrons\footnote{
Electrons are included in the calculations in
the following order:
$6p,\allowbreak 6s;\allowbreak 5d;\allowbreak 5p;\allowbreak
5s;\allowbreak 4f;\ldots$
}
using the equivalent basis sets.  One can see that the 2e-CI
(table~\ref{CI_small}) gives only rough description of the correlation
structure of mercury atom, the errors in energies of excitations from
the
ground state are 3000--6000~cm$^{-1}$ in respect to the experimental
data
or the most elaborated calculations in table~\ref{MBPT_34}. Such a large

deviation is mainly due to neglect of the OC-V correlations involving
the
$5d$ shell. Their consideration (in the 12e-CI/MBPT2,
table~\ref{MBPT_1218}) reduces these errors to a level of about
1000~cm$^{-1}$. The contributions from the correlations with the $5p$
and
$5s$ shells (the 18e-CI/MBPT2 and 20e-CI/MBPT2, tables~\ref{MBPT_1218}
and
\ref{MBPT_20}) to the excitation energies are 400--600~cm$^{-1}$ and
about
-300~cm$^{-1}$, correspondingly.  However, these contributions are
mainly
compensated and the total contribution from the correlations with both
these
shells is only 100--300~cm$^{-1}$. The correlations with the $4f$
shell (the 34e-CI/MBPT2, table~\ref{MBPT_34}) give an essential
contribution that is 600--700~cm$^{-1}$.  Our final all-electron results
for
transitions between the first four states are within 200~cm$^{-1}$ with
the experimental data~\cite{Moore} and the Relativistic Coupled Cluster
(RCC) calculation results of Eliav {\it et al}.~\cite{Eliav}. A
relatively
large deviation for the last state is rather due to the MBPT2
approximation than due to basis set incompleteness or correlations in
more
inner shells. Our estimates show that the contribution from the
correlations with the $4d$ shell to the excitation energies is of
order of 100~cm$^{-1}$.

When only the V-V correlations are considered (tables~\ref{CI_large} and

\ref{CI_small}), the GRECP errors in reproducting the all-electron
results are within 30~cm$^{-1}$. This is a good confirmation to our
previous estimates~\cite{Mosyagin} that the GRECP describes the
electronic structure in the valence region with a high accuracy.
However,
the consideration of the correlations in the outercore region
(tables~\ref{MBPT_1218} and \ref{MBPT_20}) leads to an increase of these

errors (due to a rather large smoothing region for the nodeless
outercore
$5s,\allowbreak 5p,\allowbreak 5d$ pseudospinors\footnote{
We suggest that these GRECP errors
can be seriously reduced due to more artificial
generator configuration selection and smoothing procedure
and with the help of corrections to the GRECP operator
(see~\cite{Titov}, subsection~4.5)
because these errors arise from electronic structure reorganization in
the outercore region.
})
up to 200~cm$^{-1}$. These results are in a good agreement with that
from
table~III in paper~\cite{Mosyagin} where a significant increase in the
GRECP errors can be observed for the case of excitations from the $5d$
shell.  One should expect that the contribution from the correlations in

the outercore region to low excitation energies and GRECP errors in
their
reproducting will be reduced when passing from Hg toward Rn.

The errors of different RECPs for the case of 20 correlated electrons
are
presented in table~\ref{MBPT_20}. The errors of the GRECP are only about
4
times less than that of Ross {\it et al.}'s RECP~\cite{Ross} and about 7

times less than that of the energy-adjusted PP~\cite{Hausser} for the
same
number of electrons explicitly included in calculations.  This is also
in
agreement
with the results from table~III in paper~\cite{Mosyagin} where the
ratio between GRECP errors and errors of the other RECPs was reduced
from one order of magnitude (when excitations only for the valence
electrons were under consideration) to 1.5--2 times (for the case of
excitations from the outercore shells described by nodeless
pseudospinors).
In fact, additional changes in the RECP
errors when the OC-V correlations are taken into account
(tables~\ref{CI_small} and \ref{MBPT_20}) are within 200~cm$^{-1}$ for
the
used GRECP variants and within 400~cm$^{-1}$ for the other RECPs. The
main
reason is the spinor smoothing which has similar features for all these
RECPs.

The data from the MRCI calculations with the energy-adjusted PP using
the
CIPSO method for the transitions between the first four states from
table~6 in paper~\cite{Hausser} are within 100~cm$^{-1}$ with the
experimental data but the 20 electron PP does not take into account
contributions from the correlations with the $4f$ shell (which are up to

700~cm$^{-1}$) and the used basis set does not contain functions with
$h$
orbital momentum (their contributions are up to 300~cm$^{-1}$).
Therefore,
these data are results from cancelation of a few contributions: PP
errors
(e.g., the $6s^1_{1/2}6p^1_{1/2}(J=0)$--$6s^1_{1/2}6p^1_{3/2}(J=2)$
splitting is overestimated about 1000 cm$^{-1}$ by the energy-adjusted
PP
because of the features of the spin-orbit simulation in the $LS$-based
variant
of the energy-adjusted scheme), a neglect of the correlations with the
$4f$
shell, a basis set incompleteness, etc. The estimation of error for
these
data within 5\%~\cite{Hausser} appears to be correct. The data in
paper~\cite{Alekseyev} were obtained with the 12 electron RECP of Ross
{\it et al}.\ which differs from the used here 20 electron RECP.

One can see from tables~\ref{CI_large}, \ref{CI_small}, and
\ref{MBPT_20}
that the errors of the 20 and 34 electron GRECPs are of the same order
of
magnitude because the main distinctions between these GRECP variants are

inclusion of the $4f$ electrons in calculations with the 34 electron
GRECP
and smoothing the $5f$ spinors for the 20 electron GRECP. However, the
34
electron GRECP allows one to take into account explicitly the
correlations
with the $4f$ shell which are important for an agreement with the
experimental data within 200~cm$^{-1}$. As one can see from
table~\ref{MBPT_34}, the 34 electron GRECP errors are within
200~cm$^{-1}$
in this case.

The basis set truncation effect at different orbital quantum numbers on
energies of excitations from the ground state can be observed in
tables~\ref{MBPT_34} and \ref{MBPT_gfd} for the 34e-CI/MBPT2
calculations.
It is clear that $(spd)$ correlation basis set\footnote{ The
$[12,12,11,1]$
basis set contains only such functions with $f$ orbital momentum which
correspond to the $4f_{5/2}$ and $4f_{7/2}$ spinors.
}
does not allow one to take into account the correlations with the
$5d$ shell properly.
This leads to the excitation energies (table~\ref{MBPT_gfd})
rather close to that from the 2e-CI calculations.  Addition of functions

with $f$ orbital momentum is crucial for description of these
correlations
and leads to an increase in the excitation energies on
2000--4000~cm$^{-1}$. The excitation energies for the $(spdf)$ basis set
are
within 900~cm$^{-1}$ from that for the $(spdfgh)$ basis set
(tables~\ref{MBPT_34} and \ref{MBPT_gfd}). In turn, $(spdf)$ basis set
is
inadequate for description of the correlations with the $4f$ shell. The
results for this basis set are rather close to that from the
20e-CI/MBPT2
calculations in table~\ref{MBPT_20}. Addition of functions with $g$
orbital momentum is necessary for the correct description of these
correlations and gives the contribution 300--600~cm$^{-1}$ to the
excitation energies (table~\ref{MBPT_gfd}) whereas addition of $h$
functions contributes about 300~cm$^{-1}$ (table~\ref{MBPT_34}).  Our
test
calculations showed that the contribution from functions with $i$
orbital
momentum is of order 50~cm$^{-1}$.

One can see from comparison between the 6 and 7 columns in
tables~\ref{CI_large}, \ref{CI_small}, and \ref{MBPT_20} that the errors

due to the gaussian approximation of the GRECP components are
approximately one order of magnitude less than the errors of the
numerical
GRECP. As one can see from tables~\ref{CI_large}, \ref{CI_small},
\ref{MBPT_20}, and \ref{MBPT_34}, the effects of different nuclear
models
may be neglected for accuracy within 200 cm$^{-1}$.

\section{Conclusions}

The GRECP allows one to reproduce the electronic structure in the
valence
and outercore regions essentially better than the other tested RECPs for

the same number of explicitly treated electrons.

At least 34 external electrons (occupying the $4f,\allowbreak
5s,\allowbreak 5p,\allowbreak 5d,\allowbreak 6s,\allowbreak 6p$ shells)
of
mercury atom should be correlated and the one-electron basis set should
contain up to $h$ angular momentum functions in order to
obtain a reliable agreement with experimental data for low excitation
energies within 200 cm$^{-1}$ whereas the errors of the gaussian
approximation of the GRECP components and the effects of different
nuclear
models are negligible for this accuracy. However, our test calculations
show that the main contribution from the correlations with the $4f$
shell
is due to the one-electron correction from the self-energy
diagrams~\cite{Dzuba}, therefore, this contribution can be taken into
account for 20 electron GRECP with the help of the technique proposed
in~\cite{Titov} (subsection 5.2).

 \acknowledgments
This work was implemented under the financial support of the Russian
Foundation for Basic Research (N.~M.\ and A.~T.: grants N 96--03--33036
and 96--03--00069g; M.~K.: grant N 98--02--17663).

\begin{table}
\begin{center}
\caption{ All-electron and RECP correlation
energies\protect\tablenotemark[1],
          all-electron transition energies (TE) and
          absolute errors (AE) of different RECPs in their reproducting
          from the 2 electron valence CI calculations of low-lying
states of Hg
          for the $[8,8,7,6,5,4]$ basis set (in cm$^{-1}$). }
\begin{tabular}{clrrrrrrr}
               &                                           &
&       &       &       &       &20 el.\ &20 el.\ \\
 Sym-          & Leading
&\multicolumn{2}{c}{All-el.\ }

&34 el.\ &\multicolumn{2}{c}{20 el.\ }

& RECP  &energy-\\
\cline{3-4}
 metry         & conf.\                                    &finite
&point  & GRECP &\multicolumn{2}{c}{GRECP}

&of Ross&adjusted\\
\cline{6-7}
 ($J_{parity}$)&                                           &nucl.\
&nucl.\ & num.\ & num.\ &gaus.\ &et al.\tablenotemark[3]

&PP\tablenotemark[4]\\
\hline
               &
&\multicolumn{7}{c}{Correlation energies\tablenotemark[1]}\\
\cline{3-9}
 $0_g$         & $6s_{1/2}^2$                              & -5991 &
-5987 & -6019 & -6023 & -6024 & -6006 & -5990 \\
 $0_u$         & $6s_{1/2}^1 6p_{1/2}^1$                   &  -922 &
-922 &  -927 &  -927 &  -927 &  -926 &  -933 \\
 $1_u$         & $6s_{1/2}^1 6p_{1/2}^1$\tablenotemark[2]  & -6919 &
-6916 & -6936 & -6937 & -6937 & -6923 & -6735 \\
 $2_u$         & $6s_{1/2}^1 6p_{3/2}^1$                   & -1108 &
-1108 & -1111 & -1111 & -1112 & -1111 & -1116 \\
 $1_u$         & $6s_{1/2}^1 6p_{3/2}^1$\tablenotemark[2]  &   171 &
172 &   164 &   163 &   163 &   171 &    17 \\
\cline{3-9}
               &                                           &    TE &
TE &    AE &    AE &    AE &    AE &    AE \\
\cline{3-9}
 $0_g$         & $6s_{1/2}^2$                              &     0 &
0 &     0 &     0 &     0 &     0 &     0 \\
 $0_u$         & $6s_{1/2}^1 6p_{1/2}^1$                   & 31832 &
31891 &    -1 &    17 &    16 &   203 &  -687 \\
 $1_u$         & $6s_{1/2}^1 6p_{1/2}^1$\tablenotemark[2]  & 33704 &
33764 &    -3 &    17 &    16 &   215 &  -439 \\
 $2_u$         & $6s_{1/2}^1 6p_{3/2}^1$                   & 38073 &
38136 &    -8 &    18 &    17 &   248 &   360 \\
 $1_u$         & $6s_{1/2}^1 6p_{3/2}^1$\tablenotemark[2]  & 50108 &
50162 &     7 &    34 &    33 &   245 &   -68
\end{tabular}
\tablenotetext[1]{The correlation energies were calculated as a
difference between
                  the total energies from the above mentioned CI
calculations and the
                  numerical SCF calculations with the frozen core
$1s_{1/2}$--$5d_{5/2}$ spinors
                  for the corresponding terms.}
\tablenotetext[2]{The $6s_{1/2}^1 6p_{1/2}^1$ and $6s_{1/2}^1
6p_{3/2}^1$
   configurations are strongly mixed for these terms in the CI
calculations.}
\tablenotetext[3]{The RECP from Ref.~\cite{Ross}.}
\tablenotetext[4]{The PP from Ref.~\cite{Hausser} with the corrected
 $V_{so}$ by the factors $(2l+1)/2$ (M.\ Dolg, private communication).}
\label{CI_large}
\end{center}
\end{table}

\begin{table}
\begin{center}
\caption{ All-electron and RECP correlation
energies\protect\tablenotemark[1],
          all-electron transition energies (TE) and
          absolute errors (AE) of different RECPs in their reproducting
          from the 2 electron valence CI calculations of low-lying
states of Hg
          for the $[4,4,3,2,1]$ basis set (in cm$^{-1}$). }
\begin{tabular}{clrrrrrrr}
               &                                           &
&       &       &       &       &20 el.\ &20 el.\ \\
 Sym-          & Leading
&\multicolumn{2}{c}{All-el.\ }

&34 el.\ &\multicolumn{2}{c}{20 el.\ }

& RECP  &energy-\\
\cline{3-4}
 metry         & conf.\                                    &finite
&point  & GRECP &\multicolumn{2}{c}{GRECP}

&of Ross&adjusted\\
\cline{6-7}
 ($J_{parity}$)&                                           &nucl.\
&nucl.\ & num.\ & num.\ &gaus.\ &et al.\tablenotemark[3]

&PP\tablenotemark[4]\\
\hline
               &
&\multicolumn{7}{c}{Correlation energies\tablenotemark[1]}\\
\cline{3-9}
 $0_g$         & $6s_{1/2}^2$                              & -5927 &
-5923 & -5954 & -5964 & -5965 & -5941 & -5924 \\
 $0_u$         & $6s_{1/2}^1 6p_{1/2}^1$                   &  -909 &
-910 &  -914 &  -915 &  -915 &  -913 &  -919 \\
 $1_u$         & $6s_{1/2}^1 6p_{1/2}^1$\tablenotemark[2]  & -6903 &
-6900 & -6920 & -6922 & -6923 & -6907 & -6718 \\
 $2_u$         & $6s_{1/2}^1 6p_{3/2}^1$                   & -1097 &
-1097 & -1100 & -1101 & -1101 & -1100 & -1106 \\
 $1_u$         & $6s_{1/2}^1 6p_{3/2}^1$\tablenotemark[2]  &   237 &
238 &   227 &   221 &   221 &   235 &    79 \\
\cline{3-9}
               &                                           &    TE &
TE &    AE &    AE &    AE &    AE &    AE \\
\cline{3-9}
 $0_g$         & $6s_{1/2}^2$                              &     0 &
0 &     0 &     0 &     0 &     0 &     0 \\
 $0_u$         & $6s_{1/2}^1 6p_{1/2}^1$                   & 31780 &
31839 &    -1 &    22 &    21 &   202 &  -687 \\
 $1_u$         & $6s_{1/2}^1 6p_{1/2}^1$\tablenotemark[2]  & 33655 &
33716 &    -4 &    21 &    21 &   214 &  -438 \\
 $2_u$         & $6s_{1/2}^1 6p_{3/2}^1$                   & 38020 &
38083 &    -9 &    22 &    21 &   247 &   358 \\
 $1_u$         & $6s_{1/2}^1 6p_{3/2}^1$\tablenotemark[2]  & 50109 &
50164 &     3 &    31 &    31 &   241 &   -73
\end{tabular}
\tablenotetext[1]{The correlation energies were calculated as a
difference between
                  the total energies from the above mentioned CI
calculations and the
                  numerical SCF calculations with the frozen core
$1s_{1/2}$--$5d_{5/2}$ spinors
                  for the corresponding terms.}
\tablenotetext[2]{The $6s_{1/2}^1 6p_{1/2}^1$ and $6s_{1/2}^1
6p_{3/2}^1$
   configurations are strongly mixed for these terms in the CI
calculations.}
\tablenotetext[3]{The RECP from Ref.~\cite{Ross}.}
\tablenotetext[4]{The PP from Ref.~\cite{Hausser} with the corrected
 $V_{so}$ by the factors $(2l+1)/2$ (M.\ Dolg, private communication).}
\label{CI_small}
\end{center}
\end{table}

\begin{table}
\begin{center}
\caption{ All-electron transition energies (TE) and
          absolute errors (AE) of the 20 electron GRECP in their
reproducting
          from the 12 and 18 electron CI/MBPT2 calculations
          of low-lying states of Hg
          for the $[4,4,3,2,1]$ CI, $[11,11,11,9,8,7]$ and
$[11,12,11,9,8,7]$ MBPT basis sets (in cm$^{-1}$). }
\begin{tabular}{clrrrr}
 Sym-          &                                           &All-el.\
                                                                   &20
el.\ &All-el.\

&20 el.\ \\
 metry         & Leading                                   &point  &
GRECP &point  & GRECP \\
 ($J_{parity}$)& conf.\                                    &nucl.\
&gaus.\ &nucl.\ &gaus.\ \\
\hline
\multicolumn{2}{l}{Number of correlated electrons:}        &    12 &
12 &    18 &    18 \\
\hline
               &                                           &    TE &
AE &    TE &    AE \\
\cline{3-6}
 $0_g$         & $6s_{1/2}^2$                              &     0 &
0 &     0 &     0 \\
 $0_u$         & $6s_{1/2}^1 6p_{1/2}^1$                   & 36753 &
-194 & 37264 &  -211 \\
 $1_u$         & $6s_{1/2}^1 6p_{1/2}^1$                   & 38526 &
-178 & 39049 &  -192 \\
 $2_u$         & $6s_{1/2}^1 6p_{3/2}^1$                   & 43149 &
-167 & 43751 &  -172 \\
 $1_u$         & $6s_{1/2}^1 6p_{3/2}^1$                   & 52919 &
-30 & 53325 &   -24
\end{tabular}
\label{MBPT_1218}
\end{center}
\end{table}

\begin{table}
\begin{center}
\caption{ All-electron transition energies (TE) and
          absolute errors (AE) of different RECPs in their reproducting
          from the 20 electron CI/MBPT2 calculations
          of low-lying states of Hg
          for the $[4,4,3,2,1]$ CI and $[12,12,11,9,8,7]$ MBPT basis
sets (in cm$^{-1}$). }
\begin{tabular}{clrrrrrrr}
               &                                           &
&       &       &       &       &20 el.\ &20 el.\ \\
 Sym-          & Leading
&\multicolumn{2}{c}{All-el.\ }

&34 el.\ &\multicolumn{2}{c}{20 el.\ }

& RECP  &energy-\\
\cline{3-4}
 metry         & conf.\                                    &finite
&point  & GRECP &\multicolumn{2}{c}{GRECP}

&of Ross&adjusted\\
\cline{6-7}
 ($J_{parity}$)&                                           &nucl.\
&nucl.\ & num.\ & num.\ &gaus.\ &et al.\tablenotemark[1]

&PP\tablenotemark[2]\\
\hline
               &                                           &    TE &
TE &    AE &    AE &    AE &    AE &    AE \\
\cline{3-9}
 $0_g$         & $6s_{1/2}^2$                              &     0 &
0 &     0 &     0 &     0 &     0 &     0 \\
 $0_u$         & $6s_{1/2}^1 6p_{1/2}^1$                   & 36900 &
36963 &  -137 &  -135 &  -130 &   570 &  -468 \\
 $1_u$         & $6s_{1/2}^1 6p_{1/2}^1$                   & 38690 &
38754 &  -128 &  -115 &  -111 &   581 &  -233 \\
 $2_u$         & $6s_{1/2}^1 6p_{3/2}^1$                   & 43415 &
43481 &  -127 &   -94 &   -90 &   656 &   795 \\
 $1_u$         & $6s_{1/2}^1 6p_{3/2}^1$                   & 53004 &
53062 &   -22 &    50 &    52 &   545 &   303
\end{tabular}
\tablenotetext[1]{The RECP from Ref.~\cite{Ross}.}
\tablenotetext[2]{The PP from Ref.~\cite{Hausser} with the corrected
 $V_{so}$ by the factors $(2l+1)/2$ (M.\ Dolg, private communication).}
\label{MBPT_20}
\end{center}
\end{table}

\begin{table}
\begin{center}
\caption{ All-electron transition energies (TE) and
          absolute errors (AE) of the 34 electron GRECP in their
reproducting
          from the 34 electron CI/MBPT2 calculations
          of low-lying states of Hg
          for the $[4,4,3,2,1]$ CI and $[12,12,11,10,8,7]$ MBPT basis
sets
          in comparison with experimental data and results of RCC
calculations (in cm$^{-1}$). }
\begin{tabular}{clrrrrr}
               &                                           &
&RCC\tablenotemark[2]

&       &       &       \\
 Sym-          & Leading                                   &Exper.\

&(all-el.,

&\multicolumn{2}{c}{All-el.\ }

&34 el.\ \\
\cline{5-6}
 metry         & conf.\
&data\tablenotemark[1]

&finite &finite &point  & GRECP \\
 ($J_{parity}$)&                                           &
&nucl.) &nucl.\ &nucl.\ & num.\ \\
\hline
               &                                           &    TE &
TE &    TE &    TE &    AE \\
\cline{3-7}
 $0_g$         & $6s_{1/2}^2$                              &     0 &
0 &     0 &     0 &     0 \\
 $0_u$         & $6s_{1/2}^1 6p_{1/2}^1$                   & 37645 &
37453 & 37569 & 37634 &   153 \\
 $1_u$         & $6s_{1/2}^1 6p_{1/2}^1$                   & 39412 &
39302 & 39361 & 39426 &   169 \\
 $2_u$         & $6s_{1/2}^1 6p_{3/2}^1$                   & 44043 &
44190 & 44157 & 44224 &   225 \\
 $1_u$         & $6s_{1/2}^1 6p_{3/2}^1$                   & 54069 &
55453 & 53553 & 53612 &   224
\end{tabular}
\tablenotetext[1]{The data from Ref.~\cite{Moore}.}
\tablenotetext[2]{The results from Ref.~\cite{Eliav} for the
$[27,23,21,16,10,6]$ basis set (in the notations of the present paper)
and 34 correlated electrons.}
\label{MBPT_34}
\end{center}
\end{table}

\begin{table}
\begin{center}
\caption{ All-electron transition energies (TE)
          from the 34 electron CI/MBPT2 calculations
          of low-lying states of Hg
          for different basis sets (in cm$^{-1}$). }
\begin{tabular}{clrrr}
 Sym-          &                                           &All-el.\

&All-el.\

&All-el.\ \\
 metry         & Leading                                   &point
&point  &point  \\
 ($J_{parity}$)& conf.\                                    &nucl.\
&nucl.\ &nucl.\ \\
\hline
\multicolumn{2}{l}{CI basis:}
&$[4,4,3,2,1]$

&$[4,4,3,2]$

&$[4,4,3]$\\
\multicolumn{2}{l}{MBPT basis:}
&$[12,12,11,9,8]$

&$[12,12,11,9]$

&$[12,12,11,1]$\\
\hline
               &                                           & TE    &
TE    & TE    \\
\cline{3-5}
 $0_g$         & $6s_{1/2}^2$                              &     0 &
0 &     0 \\
 $0_u$         & $6s_{1/2}^1 6p_{1/2}^1$                   & 37370 &
37068 & 32680 \\
 $1_u$         & $6s_{1/2}^1 6p_{1/2}^1$                   & 39151 &
38798 & 34530 \\
 $2_u$         & $6s_{1/2}^1 6p_{3/2}^1$                   & 43906 &
43296 & 38847 \\
 $1_u$         & $6s_{1/2}^1 6p_{3/2}^1$                   & 53272 &
52726 & 50604
\end{tabular}
\label{MBPT_gfd}
\end{center}
\end{table}


\begin{references}
\bibitem[*]{Email}      E-mail: Mosyagin@hep486.pnpi.spb.ru
                        for correspondence; \hfill ~ \linebreak
                        http://www.qchem.pnpi.spb.ru
\bibitem{Papers}
                        E.~Eliav, U.~Kaldor, Y.~Ishikawa, M.~Seth,
                        and P.~Pyykko,
                        Phys.\ Rev.\ A {\bf 53}, 3926 (1996);
                        F.~Rakowitz and C.~M.~Marian,
                        Chem.\ Phys.\ Lett.\ {\bf 257}, 105 (1996);
                        U.~Wahlgren, M.~Sjovoll, H.~Fagerli, O.~Gropen,
                        and B.~Schimmelpfenning,
                        Theor.\ Chim.\ Acc.\ {\bf 97}, 324 (1997);
                        R.~J.~Buenker, A.~B.~Alekseyev,
H.-P.~Liebermann,
                        R.~Lingott, and G.~Hirsch,
                        Phys.\ Rev.\ A, in press.
\bibitem{Dzuba}         V.~A.~Dzuba, V.~V.~Flambaum, and M.~G.~Kozlov,
                        Pis'ma v ZheTF {\bf 63}, 844 (1996);
                        V.~A.~Dzuba, V.~V.~Flambaum, and M.~G.~Kozlov,
                        Phys.\ Rev.\ A {\bf 54}, 3948 (1996);
                        M.~G.~Kozlov and S.~G.~Porsev,
                        ZheTF {\bf 111}, 838 (1997)
                        [in Russian].
\bibitem{Eliav}         E.~Eliav, U.~Kaldor, and Y.~Ishikawa,
                        Phys.\ Rev.\ A {\bf 52}, 2765 (1995).
\bibitem{Mosyagin}      N.~S.~Mosyagin, A.~V.~Titov, and Z.~Latajka,
                        Int.\ J.\ Quant.\ Chem.\ {\bf 63}, 1107 (1997).
\bibitem{Titov}         A.~V.~Titov and N.~S.~Mosyagin,
                        Preprint PNPI No.\ {\bf 2182} (Petersburg
Nuclear
                        Physics Institute, Gatchina, St.-Petersburg
district, 1997),
                        81~p., submitted for publication.
\bibitem{Ross}          R.~B.~Ross, J.~M.~Powers, T.~Atashroo,
W.~C.~Ermler,
                        L.~A.~Lajohn, and P.~A.~Christiansen,
                        J.\ Chem.\ Phys.\ {\bf 93}, 6654 (1990).
\bibitem{Hausser}       U.~Haussermann, M.~Dolg, H.~Stoll, H.~Preuss,
                        P.~Schwerdtfeger, and R.~M.~Pitzer,
                        Mol.\ Phys.\ {\bf 78}, 1211 (1993).
\bibitem{Tupitsyn}      I.~I.~Tupitsyn, N.~S.~Mosyagin, and A.~V.~Titov,

                        J.\ Chem.\ Phys.\ {\bf 103}, 6548 (1995).
\bibitem{Titov1}        A.~V.~Titov, A.~O.~Mitrushenkov, and
I.~I.~Tupitsyn,
                        Chem.\ Phys.\ Lett.\ {\bf 185}, 330 (1991).
\bibitem{Moore}         C.~E.~Moore,
                        Circ.\ Natl.\ Bur.\ Stand.\ (U.S.) {\bf 467}
(1958).
\bibitem{Alekseyev}     A.~B.~Alekseyev, H.-P.~Liebermann,
R.~J.~Buenker,
                        and G.~Hirsch,
                        J.\ Chem.\ Phys.\ {\bf 104}, 4672 (1996).
\end{references}
\end{document}